\documentstyle [prl,aps]{revtex}
\begin{document}
\twocolumn[{
\widetext
\draft

\title{
Destruction of density-wave states by
a pseudo-gap in high magnetic fields: application to
(TMTSF)$_2$ClO$_4$
}

\author{Ross H. McKenzie\cite{email}}

\address{School of Physics, University of New
South Wales, Sydney, NSW 2052, Australia}

\date{December 23, 1994}
\maketitle
\mediumtext
\begin{abstract}
A model is presented for the
destruction of density-wave states in
quasi-one-dimensional crystals
by high magnetic fields. The model is consistent with
previously unexplained properties of the organic conductors
(TMTSF)$_2$ClO$_4$ and
(BEDT-TTF)$_2$MHg(SCN)$_4$ (M=K,Rb,Tl).
As the magnetic field increases quasi-one-dimensional
density-wave fluctuations increase, producing
a pseudo-gap in the electronic density of states near
the transition temperature.
When the pseudo-gap becomes larger than the mean-field
transition temperature formation of a density-wave state
is not possible.
\\

Accepted for publication in {\it Physical Review Letters} \\
\end{abstract}

%75.30.Fv Spin-density waves
%71.45.Lr Charge-density-wave systems
%74.70.Kn Organic superconductors
%75.30.Kz Magnetic phase boundaries (including magnetic transitions,
%         metamagnetism, etc.)
%68.35.Rh Phase transitions and critical phenomena
%72.15.Gd Galvanomagnetic and other magnetotransport effects

}]
\narrowtext

Understanding low-dimensional electronic systems continues to
be one of the major challenges of
condensed matter physics.
Organic conductors are particulary interesting systems
because they exhibit a subtle competition between
metallic, superconducting, charge-density-wave (CDW)
and spin-density-wave (SDW) phases. This competition
is sensitive to pressure, temperature, magnetic
field, and chemical substitution \cite{ish,bro,kan}.
This Letter presents a theory of the
destruction of density-wave phases in quasi-one-dimensional
materials in high magnetic fields.
The model is used to understand previously unexplained
properties of the quasi-one-dimensional conductor
(TMTSF)$_2$ClO$_4$
and the quasi-two-dimensional conductors
(BEDT-TTF)$_2$MHg(SCN)$_4$ (M=K,Rb,Tl).

At ambient pressure and zero magnetic field
(TMTSF)$_2$ClO$_4$ is a superconductor below 1.3 K.
A spectacular cascade of transitions
into field-induced-spin-density-wave (FISDW) phases
occurs at fields above 4 T. The quantum Hall
effect is observed in these phases \cite{ish}.
The phase-diagram below 15 T can be explained in
terms of the so-called ``standard model'' which is
a mean-field treatment of a highly-anisotropic
Hubbard model \cite{ish,gor}.
However, there are at least five experimental observations
concerning the behaviour of (TMTSF)$_2$ClO$_4$ in
high magnetic fields that cannot be explained by any existing theory.
(a) At about 15 T the field-induced spin-density-wave transition
temperature $T_{SDW}$ reaches a maximum of about 5 K
and then decreases at higher fields, going to
zero at 27 T \cite{nau}. The high field
phase is known as the ``re-entrant phase.''
Its existence contradicts the standard model in which
 the transition temperature saturates at high fields.
Specific heat measurements indicate that there
is a well-defined phase transition from the FISDW
phase to the re-entrant phase \cite{for,for2}.
(b) Below (above) 2 K  the specific heat
 of the re-entrant phase at 30 T is smaller (larger)
than that of the metallic phase at zero-field \cite{for,for2}.
(c) Along the phase boundary at high temperatures
the ratio of the specific
heat jump, $\Delta C$, to the normal state electronic
specific heat, $\gamma T$, is larger than the mean-field
value of 1.43. For example, at 15 T this ratio is about
 3.5 \cite{for,for2}.
On the phase boundary near 27 T the ratio is less than
the BCS value.
(d) In the re-entrant phase the magnetoresistance is approximately
activated in temperature \cite{nau}.
(e) Thermopower measurements suggest that there is an electronic
energy gap above 30 T and below 2.6 K\cite{yu}.

Previously two models have been proposed
to explain the existence of the re-entrant phase \cite{yak,ani}.
Yakovenko\cite{yak} considered how a high field
could effectively confine the electronic motion
to single stacks of TMTSF molecules and as a result
the FISDW phase would be destroyed by one-dimensional
fluctuations. However, this model
does not predict the observed rapid decrease
of $T_{SDW}$ with field \cite{for} and if commonly accepted
parameter values are used the
theory will only be relevant for fields of the order of 100 T \cite{cha2}.
The second model\cite{ani} involves changes in the band structure
produced by the ordering of the ClO$_4$ ions at 24 K.
This model predicts that there should be spikes in the
transition temperature, periodic in the inverse field, at high fields.
Although there is some evidence for such spikes in acoustic
 experiments\cite{shi}
there is no evidence for such  spikes in specific heat,\cite{for,for2}
 resistivity,\cite{nau}
and thermopower\cite{yu} measurements.
This model also
does not predict the observed rapid decrease
of $T_{SDW}$ with field \cite{for,for2}.

This Letter presents a model for behaviour near
the FISDW-re-entrant phase boundary that is
consistent with the five observations listed above.
Before describing the details a brief description
is given  of the basic physics.
A SDW forms when the opening of an energy gap
at the Fermi surface, due to the SDW, lowers the
total electronic energy by more than the increase in
Coulomb energy due to the SDW. The size of the
energy gap is proportional to the amplitude
of the SDW. However, the fluctuations
in the SDW order parameter affect the
electronic states.
Similar effects have been seen in CDW systems
\cite{mck,lon,dre}.
Such effects are ignored in the standard model.
Near the transition temperature
the fluctuations increase and
there are  long range SDW correlations
producing a pseudo-gap in the density of states.
Furthermore, the pseudo-gap reduces the transition temperature
and for a sufficiently large pseudo-gap
formation of a SDW is not possible (Figure \ref{figpsgap}).
The reason for this is simple. In the presence of a large
pseudo-gap, opening an energy gap due to a SDW
will not lower the total energy sufficiently to make
formation of a SDW energetically favourable.
The size of the pseudo-gap is
determined by the magnitude of the SDW fluctuations
 which are in turn affected by the
size of the SDW correlations {\it transverse}
to the chains.  Bjeli\v{s} and Maki  \cite{bje}
have shown that the transverse correlation
length is a decreasing function of magnetic field.
This provides a mechanism for the FISDW-re-entrant
phase transition seen in (TMTSF)$_2$ClO$_4$:
as the field increases the transverse correlation length decreases
and the pseudo-gap increases above a critical value.
The presence of a pseudo-gap is consistent
with the observations (b), (d) and (e).

It is generally believed that the physics of the
SDW phases found in the (TMTSF)$_2$X
salts can be described by a Hubbard model
with highly anisotropic dispersion \cite{ish}.
The hopping integrals $t_a$, $t_b$, and $t_c$
associated with the three crystal axes are estimated
to have values of about 250 meV, 25 meV, and 1 meV,
respectively.
The Fermi surface consists of two slightly warped planes.
The sensitivity of the {\it imperfect nesting} of
this open Fermi surface to pressure, magnetic field,
and the anion X, is responsible
for the rich phase diagram.
A mean-field treatment of the anisotropic Hubbard model
 can explain the existence of the
FISDW phases \cite{ish,gor}. At zero-field the imperfect nesting
prevents formation of a SDW. A magnetic field improves
the nesting because the electron motion becomes
more one dimensional \cite{gor,cha} resulting in SDW formation
and a maximum transition temperature of about 5 K at 15 T.
Since, only behaviour above 15 T is considered here,
for simplicity, perfect nesting is assumed and
a one-dimensional model of the electronic states
is considered.
Consequently, the effect of the orbital electron
motion on the density of states and the transition temperature
is neglected.

At the mean-field level the electrons
move in a potential with wave vector $2k_F$, $\Delta(x)$,
 that is proportional to the SDW amplitude.
The upper and
lower components of a spinor $\Psi(x)$ describe left-moving, up-spin and
 right-moving, down-spin electrons,
respectively, with Fermi velocity $v_F$.
The Hamiltonian for these electrons is \cite{ish}
\begin{equation}
H = \int dx \Psi^\dagger \bigg[ - iv_F \sigma_3
{\partial \over \partial x} + {1 \over 2}(\Delta(x) \sigma_+ + h.c.
)\bigg] \Psi
\label{hamel}
\end{equation}
where $\sigma_3$ and
$ \sigma_{\pm} \equiv \sigma_1 + i \sigma_2$ are Pauli matrices.
It is assumed that the SDW is incommensurate with the
lattice and so $\Delta(x)$ is complex.
The other  electrons are described by a similar Hamiltonian.

In the standard mean-field treatment
the order parameter $\Delta(x)$
is replaced by its expectation value.
In reality $\Delta(x)$ is a dynamical field
that fluctuates due to quantum and thermal effects.
These fluctuations are key to the model presented here.
Such spin excitations
have been detected in antiferromagnetic
resonance experiments \cite{tor} and  have been used to
explain the temperature dependence of the SDW amplitude \cite{le}.
The characteristic energy scale of these excitations is about 1 K
and they soften as the transition temperature is approached.
Consequently, these excitations are treated classically here.
At the Gaussian level the SDW correlation function
above the transition temperature is \cite{sch}
\begin{equation}
\langle \Delta(x)\Delta(x')^* \rangle = \psi^2
\exp(-|x-x'|/\xi_a(T)).
\label{cor2}
\end{equation}
where $\xi_a(T)$ is the correlation length along the
chains and $\psi$ is the rms fluctuation in the
order parameter. For a strictly one
dimensional system
$\psi^2 \propto T \xi_a(T) \to \infty$
close to a phase transition \cite{sch}.
(This divergence  is related to the fact that in one dimension fluctuations
prevent finite temperature phase transitions).
A more realistic model takes into account the
coupling between chains and
with  $q_0$ a wavevector cutoff gives
\begin{equation}
 \psi^2 \propto q_0 T
{  \xi_a(T)^2 \over \xi_b(T)\xi_c(T)}
\label{cf}
\end{equation}
which is finite as $ \xi_i(T) \to \infty$.
Here $\psi$ will be treated as a parameter
that is a measure of the SDW fluctuations.
The important point is that {\it as the transverse correlation length $\xi_b$
decreases $\psi$ increases.}

Bjeli\v{s} and Maki \cite{bje}
considered the effect of a large magnetic field
parallel to the c direction (the least conducting direction)
on the CDW and SDW correlations length in the a and b directions.
They showed that, while $\xi_a$ was not significantly affected,
$\xi_b$ is a decreasing function of magnetic field
(inset of Figure \ref{figpsgap}).
This is because as
as the field increases the electron   motion
 becomes more one-dimensional \cite{gor,cha}.
If $\omega_c$ is the cyclotron frequency and $T_{MF}$
is the mean-field transition tempearature then the
size of the reduction is determined  by the ratio
\begin{equation}
{\omega_c \over T}
\equiv {eb v_F B \over T}
\equiv { B \over B_0}
{ T_{MF} \over T}
\label{aa1}
\end{equation}
For (TMTSF)$_2$ClO$_4$
 ($b=7.7 \AA$, $v_F=2 \times 10^5$ m/sec, and $T_{MF}= 5$ K)
$B_0 \sim 3$ T \cite{kri}.

This Letter uses the following model for the fluctuations
in $\Delta(x)$. It
is replaced in (1) with a {\it static} random potential with zero mean,
$\langle\Delta(x)\rangle = 0$,
 and correlations given by (\ref{cor2}).
Treating $\Delta(x)$ as a {\it static} field
is a reasonable approximation if the fluctuations
can be treated classically.
Similar arguments have been used to successfully
model the effect of lattice fluctuations on
the electronic properties of CDW compounds \cite{mck}.

Sadovsk\~i\~i \cite{sad2}
 calculated the one-electron Green's function
 for the one-dimensional model (\ref{hamel})
and (\ref{cor2}) {\it exactly.}
A perturbative treatment of this problem was
given earlier \cite{ric}.
Sadovsk\~i\~i found that the Green's function reduced to a simple analytic form
in the limit of large correlation lengths ($\xi_a \gg v_F/\psi$)
\cite{sad}.
Since we are interested in behaviour near $T_{SDW}$
we will also take this limit.
Then it is also possible to evaluate exactly higher-order
Green's functions such as those needed to find the
transition temperature.
The density of states, shown in the inset of Figure \ref{figspec},
is zero at the Fermi energy and suppressed on an energy scale
of order $\psi$, i.e., there is a pseudo-gap.
As $\xi_a \psi/v_F$ decreases the pseudo-gap gradually fills in
\cite{sad2}.
In some CDW systems, at zero field,
optical and susceptibility measurements near the transition temperature
are consistent with a pseudo-gap due to fluctuations
\cite{dre}.

$T_{SDW}/T_{MF}$ is a universal function of $\psi/T_{MF}$
(Figure \ref{figpsgap}).
(Identical results are obtained for CDW's if Zeeman splitting is
neglected).
As the fluctuations increase $T_{SDW}$
decreases. The most important point is:
{\it for $\psi > T_{MF}$ formation of a SDW
is not possible.}

This model is consistent with the five properties
of (TMTSF)$_2$ClO$_4$ listed above.
(a) It is postulated that the FISDW-re-entrant transition
is due to destruction of the FISDW phase by a pseudo-gap.
As the field increases $\psi $ increases
 due to increasing anisotropy
 and when $\psi \sim T_{MF}$ the re-entrant transition occurs.

(b) Figure \ref{figspec} shows the temperature dependence of
the electronic specific heat
in the presence of the pseudo-gap.
 At low (high) temperatures the specific
heat is less (more) than the value in the absence of the
pseudo-gap. For comparison the observed temperature dependence
at 30 T\cite{for2} is also shown.  The data is consistent with
the requirement of the model that $\psi \sim T_{MF}\sim $ 5 K.

(c) Calculation of the specific heat jump at the phase boundary
requires knowledge of the temperature dependence of the
pseudo-gap. The results for a simple model, based on
equation (\ref{cf}), are shown in
Figure \ref{figspec}.  The specific heat jump
is significantly enhanced over the mean-field value of $1.43 \gamma T$.
This is consistent with the observed behaviour at high temperatures,
but inconsistent with the observed behaviour at low temperatures.
This disagreement could be because of the simplistic model used
for the temperature dependence of $\psi$ and because
 the SDW fluctuations can no longer
be treated classically.

(d) Due to the pseudo-gap the magnetoresistance will be approximately
activated in temperature.

(e) The thermopower data is just as consistent with a pseudogap
as an absolute gap.

Several key experiments could test this model.
Far-infra-red or nmr measurements could reveal the
pseudo-gap near the phase boundary. As the field
increases above 27 T, $\xi_a$ decreases, and the
pseudo-gap will fill in \cite{sad2}.
The anion gap model \cite{ani} predicts that the
re-entrant phase only exists due to the anion
ordering in (TMTSF)$_2$ClO$_4.$
In contrast the model presented here predicts the
destruction of SDW phases in {\it any} material at
sufficiently high fields. The difference between the two
models could be tested
by searching for the re-entrant phase above 30 T in several materials.
(i) (TMTSF)$_2$ClO$_4$: at pressures above 5 kbar
the anion ordering is destroyed \cite{kan}.
(ii) (TMTSF)$_2$PF$_6$ has no anion ordering. At ambient pressure
 $T_{SDW} \sim $ 12 K up to 30 T.
Some of these experiments are planned at the
Australian National Pulsed Magnet Laboratory
which provides access to fields up to 60 T at
temperatures down to 60 mK.

This theory is also applicable to the quasi-two-dimensional
materials (BEDT-TTF)$_2$MHg(SCN)$_4$ (M=K,Rb,Tl).
A co-existing quasi-one-dimensional
Fermi surface is believed to be responsible for
the formation of a density-wave state in these materials at
low temperatures.
There is some controversy as to whether this is a
CDW or SDW state \cite{bro}.
This state is destroyed above the ``kink field'', $H_k$ \cite{bro,bro2}.
(For M=Rb,  $H_k= $ 32 T).
The observed phase diagram is consistent with Figure \ref{figpsgap},
including the observation of first-order hysteretic behaviour
near $H_k$ at low temperatures \cite{bro,bro2}.

In conclusion, a theory has been presented to show
how fluctuations enhanced by a high magnetic field
can destroy density-wave states in quasi-one-dimensional
materials. The model is consistent with previously
unexplained properties of
(TMTSF)$_2$ClO$_4$ and (BEDT-TTF)$_2$MHg(SCN)$_4$ (M=K,Rb,Tl).
For calculational and conceptual simplicity,
the effects of anion ordering and imperfect nesting
of the Fermi surface are neglected.
A complete description of these materials must include the
effects of anion ordering, imperfect nesting,
and fluctuations.
Finally, I hope this work will stimulate others to use
sophisticated many-body techniques to test whether
the model presented here gives a good description of
the effects of spin-density-wave fluctuations on
the electronic properties of quasi-one-dimensional systems.

This work was stimulated by the visit of J. S. Brooks,
supported by NSF DMR 92-14889, to the
Australian National Pulsed Magnet Laboratory.
I have benefitted greatly  from discussions with Brooks.
I thank him for pointing out the relevance of this model to
(BEDT-TTF)$_2$MHg(SCN)$_4$.
Work at UNSW was supported by the Australian Research Council.
Early stages of this work were performed at
 Ohio State University
and supported by the U.S. Department of Energy,
Basic Energy Sciences, Division of Materials Science.
I thank J. W. Wilkins, R. A. Lewis, and D. J. Evans for helpful discussions.

\begin{figure}
\caption{Phase diagram for a spin-density-wave state in
the presence of a pseudo-gap $\psi$.
Both the pseudo-gap and the transition temperature are normalized
to the mean-field  transition temperature, $T_{MF}$.
Along the dotted lines the transition is first order.
There is a coexistence of phases between the dotted lines.
Above the transition temperature as the field increases
there is a smooth crossover from a normal metal
to a non-metal with a pseudo-gap.
This phase diagram is consistent with that of
(TMTSF)$_2$ClO$_4$ at fields above 15 T
and of (BEDT-TTF)$_2$MHg(SCN)$_4$ (M=K,Rb,Tl).
Inset: Reduction of the SDW correlation length transverse to the
chains, $\xi_b$, by a magnetic field parallel to the
least conducting direction \protect\cite{bje}.
Equation (\protect\ref{cf}) then implies that $\psi$
increases with field. $\omega_c$ is the
cyclotron frequency (see equation (\protect\ref{aa1})).
\label{figpsgap}}
\end{figure}

\begin{figure}
\caption{
Dependence of the normal state electronic specific heat $C_n$
and the specific heat jump at the phase boundary $\Delta C$ on
the pseudo-gap.
At low (high) temperatures $C_n$ is smaller (larger)
than its value, $\gamma T$, in the absence of the pseudo-gap.
$\Delta C$ is enhanced compared to the mean-field value of $1.43 \gamma T$.
As the pseudo-gap $\psi$ increases $\Delta C$
increases until $\psi \sim 2.6 T$ where
the transition becomes first order.
The triangles are $C_n$ data for
(TMTSF)$_2$ClO$_4$ at 30 T \protect\cite{for2},
with $\psi= T_{MF}=5$ K.
Inset: Pseudo-gap in the density of states near the transition
temperature. The energy is relative to the
Fermi energy. The density of states is normalized
to the free-electron value $\rho_o$.
\label{figspec}}
\end{figure}

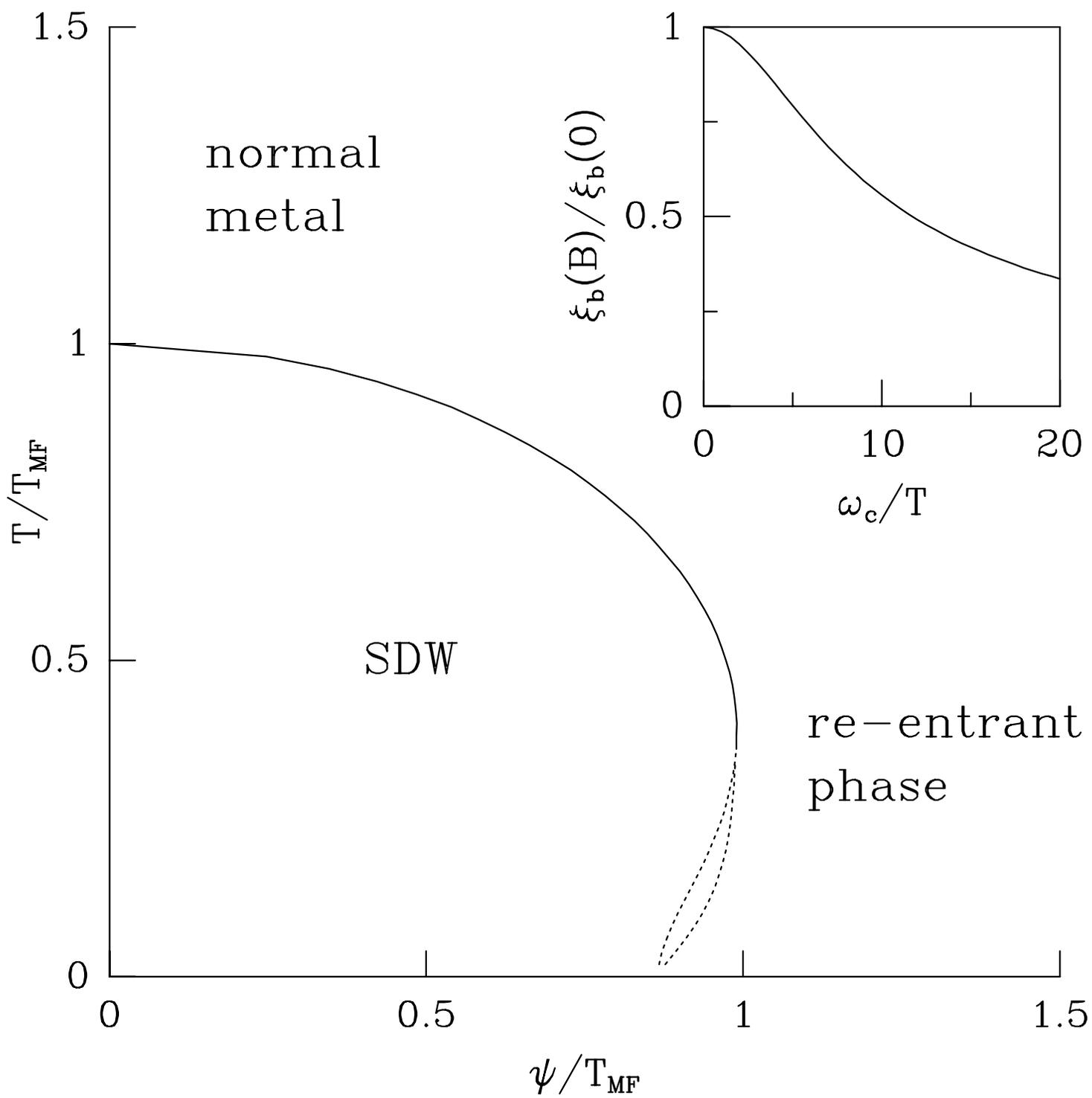
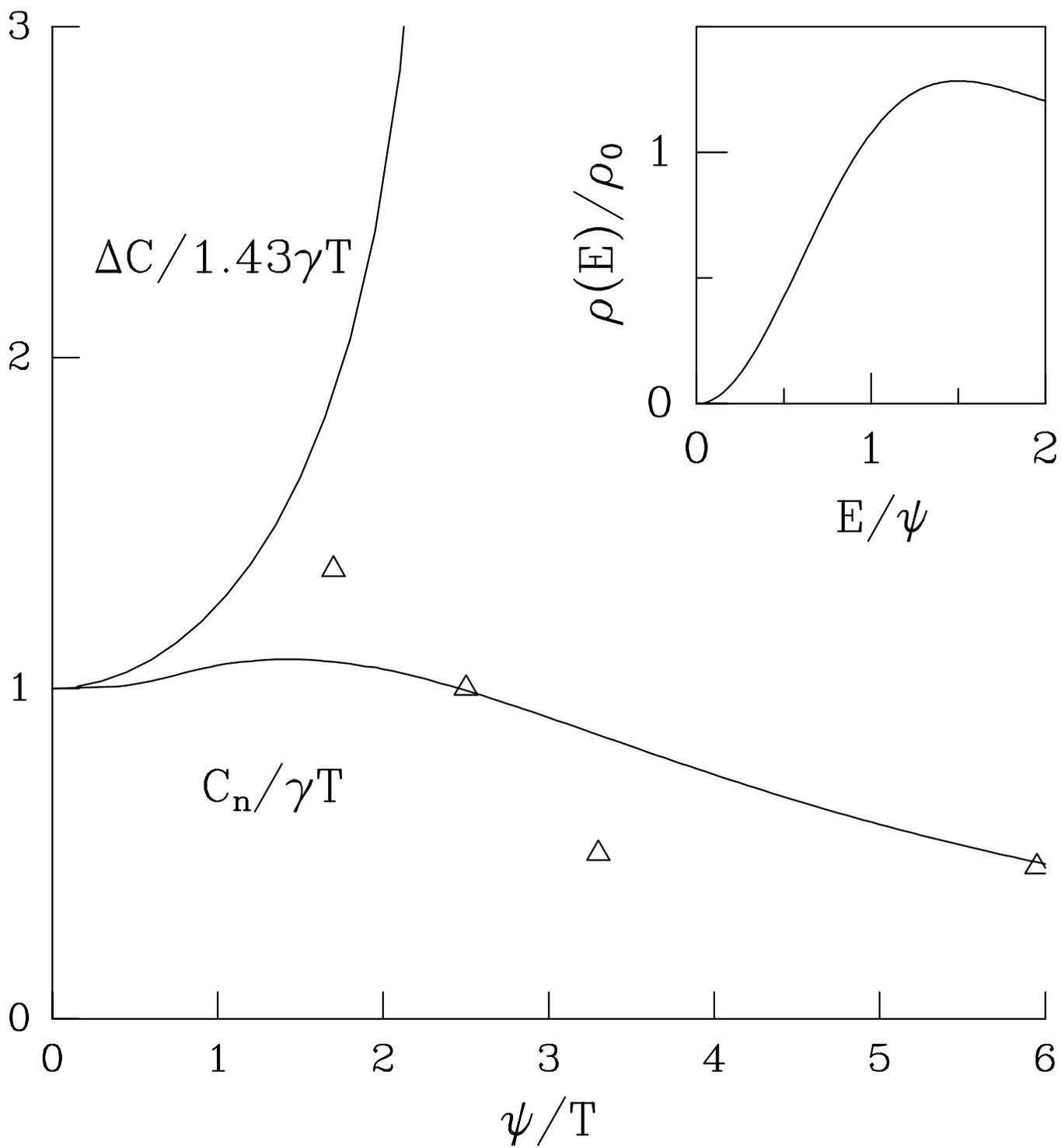
\end{document}